\documentclass[aps,prl,twocolumn,floatfix,showpacs,superscriptaddress]{revtex4}
\usepackage{graphicx}
\usepackage{ulem}
\usepackage{color}
\usepackage{bbm}
\usepackage{amsmath,amssymb}
\newcommand{\be}{\begin{equation}}
\newcommand{\beq}{\begin{equation}}
\newcommand{\ee}{\end{equation}}
\newcommand{\bea}{\begin{eqnarray}}
\newcommand{\eea}{\end{eqnarray}}
\newcommand{\ba}{\begin{array}}
\newcommand{\ea}{\end{array}}

\renewcommand{\vr} {{\bf r}}

\newcommand{\vu} {{\bf u}}

\newcommand{\vj} {{\bf j}}

\renewcommand{\vr} {{\bf r}}

\begin{document}
\title{Spin entanglement in atoms and molecules} 
\author{S. Pittalis}
\email[Electronic address:\;]{stefano.pittalis@nano.cnr.it}
\affiliation{S3 Istituto Nanoscienze, Consiglio Nazionale delle Ricerche, I-41100 Modena, Italy}
\author{F. Troiani}
\affiliation{S3 Istituto Nanoscienze, Consiglio Nazionale delle Ricerche, I-41100 Modena, Italy}
\author{C. A. Rozzi}
\affiliation{S3 Istituto Nanoscienze, Consiglio Nazionale delle Ricerche, I-41100 Modena, Italy}
\author{G. Vignale}
\affiliation{Department of Physics, University of Missouri, Columbia, Missouri 65211, USA}
\date{\today}

\begin{abstract}

We investigate the effects of inhomogeneities on spin entanglement in many-electron systems from an {\it ab-initio} approach.
The key quantity in our approach is the local spin entanglement length, which is derived from the local concurrence of the electronic system.  
Although the concurrence for an interacting systems is a highly nonlocal functional of the density, it does have a simple, albeit approximate expression in terms of Kohn-Sham orbitals. We show that the electron localization function -- well known in quantum chemistry as a descriptor of atomic shells and molecular bonds -- can be reinterpreted in terms of the ratio of the local entanglement length of the inhomogeneous system to the entanglement length of a homogenous system at the same density.
We find that the spin entanglement is remarkably enhanced in atomic shells and molecular bonds. 

\end{abstract}

 
\maketitle

{\it Introduction ---}
Entanglement arguably represents the quintessential quantum mechanical effect \cite{Schroedinger}, and has recently emerged as a crucial resource in quantum technology \cite{Nielsen}. The concept of entanglement was originally referred to the case of non-identical or spatially separated particles. 
Indistinguishable quantum particles within a many-body system cannot be labelled.  In such cases, one needs to reconsider the definition of entanglement by introducing suitable criteria for identifying the subsystems \cite{Amico08}.
In the case of a many-electron system, for example, the subsystems can be identified with the lattice sites \cite{Zanardi02} or with a set of relevant orbitals \cite{Schliemann01,Vedral03,Shi04,Boguslawski12}, and their states are defined by the corresponding  occupation numbers.  Most of the potential applications can however be related to the so-called {\it entanglement of particles}~\cite{Wiseman03,Dowling06}, where each subsystem is associated to a definite number of particles (typically one). The entanglement between two electron spins, located at two given positions within an extended many-body system \cite{Hofstetter09}, belongs to this latter class. 

The main features of such spin entanglement in an ideal Fermi gas can be derived from very general physical principles \cite{Vedral03,Oh04,Lunkes05}. In fact, the Pauli exclusion principle implies that the state of two electrons localized at the same position is necessarily a spin singlet. Besides, the spatial extension of such singlet-like character -- and of the resulting spin entanglement -- is of order of $1/k_F$ ($k_F$ being the Fermi wave vector). 
This seems to suggest that spin entanglement in real many-electron systems may be remarkably short-ranged, with a characteristic length scale that depends on the particle density.
In this Letter, we show that the situation is actually far more rich and interesting in inhomogeneous systems, such as atoms and molecules.  Here, the degree of spin  entanglement is strongly dependent not only on the distance between the electrons but also on their positions: these dependencies  can reveal the formation of atomic shells and molecular covalent bonds.   

The entanglement between two qubits can be quantified in terms of the so-called {\it concurrence} ($C$) \cite{Wootters98}, which, in our case, is a function of the positions,  $\vr_1$ and $\vr_2$, of the two electrons -- the spins of the electrons being the two qubits in question.  
Our  calculations are carried out within the framework of density functional theory (DFT).  In the Kohn-Sham  formulation, this theory replaces the actual state of the interacting many-electron system  with the state of an auxiliary noninteracting one --  the Kohn-Sham  (KS) system -- which yields the same density.  This approach provides useful insights for not too strongly correlated systems, such as atoms, and molecules far from their dissociation limit. We particularly focus on the short-range behavior of the concurrence, which is relevant when $\vr_1$ and $\vr_2$ are separated by a small displacement $\vu$.  We find that the concurrence decays as a function of $u$ over a characteristic length, the {\it entanglement length} $l_E$. In atomic and molecular systems $l_E$ is a strong function of position, which reflects the atomic shells and the molecular bond structures. Therefore, $l_E$ of, say, a molecule, greatly differs from that of a uniform Fermi gas.  In fact, the electron-localization function (ELF), often invoked in quantum chemistry as a useful tool for the visualization of atomic shells and bonds, can be reinterpreted  as a ``measure" of this difference.
From a density-functional point of view, the entanglement length is found to depend not only on the local particle-density, but also on the kinetic energy density, the gradients of the electronic density, and the paramagnetic-current density, taken in a gauge invariant combination. This implies that the concurrence is a {\it nonlocal} and {\it implicit}  functional of the density.    The rigorous structure of this functional is amenable to study by many-body theoretical methods.  
Here, we take a first fundamental step in this direction by expressing the concurrence in terms of KS orbitals.

{\it Spin entanglement in an $N$-electron system ---}
The degree of  spin entanglement between electrons localized at two different positions $\vr_1$ and $\vr_2$ can be evaluated from the spin-dependent 
two-particle reduced density matrix
\be\label{rho2}
\rho_{2} (x_1,x_2;x_1',x_2') = 
\langle 
\hat\psi^\dagger (x_2') \hat\psi^\dagger (x_1') \hat\psi (x_1) \hat\psi (x_2) 
\rangle\,,
\ee
where $x \equiv (\vr,\sigma)$ is a composite position-spin variable.  The diagonal elements of interest are obtained by taking $\vr_1=\vr_1'$ and $\vr_2=\vr_2'$. The resulting  matrix $\rho_2(\vr_1,\vr_2)$ --  with understood indices $(\sigma_1,\sigma_2)$ and   $(\sigma_1',\sigma_2')$ --  represents  
the state of a two-qubit system.  For closed-shell systems 
the reduced density matrix is diagonal in the singlet-triplet basis of the two-qubit system and has the form of a Werner state \cite{Werner89}.
Normalization with respect to spin-trace, yields the following expression: 
\begin{equation}\label{werner}
\rho^{\rm norm}_{2} (\vr_1,\vr_2) = p(\vr_1,\vr_2)\, |\Psi^-\rangle\langle\Psi^-| + \left[1-p(\vr_1,\vr_2) \right] \mathcal{I} / 4 ,
\end{equation}
where \ $\mathcal{I}$ is the identity matrix in the two-spin space and 
$
| \Psi^{-} \rangle = \frac{1}{\sqrt{2}} \left( | \uparrow\downarrow\rangle - | \downarrow\uparrow \rangle  \right)\;
$
is the singlet state.  Here $p(\vr_1,\vr_2)$, varying in the range $0<p<1$, determines the difference between the occupation of the singlet state and that of any of the three equivalent triplet states.
The concurrence of the above Werner state can be shown to be \cite{Oh04}
\begin{equation}\label{conc}
C (\vr_1,\vr_2) = \max \{[3p(\vr_1,\vr_2)-1]/2,0\}\,.
\end{equation}
The concurrence thus vanishes when the occupation of the singlet state is less than $1/2$ (i.e., for $p>1/3$), and grows monotonically for $p>1/3$, reaching the theoretical maximum $C=1$ for the singlet state ($p=1$).

As the number of electrons increases 
a direct computation of the two-particle reduced density matrix becomes rapidly 
prohibitive.  Within DFT, however, the wave function is represented by a single Slater determinant,
which satisfies the antisymmetry condition and yields, in principle, the exact ground-state
density of the interacting system.  Such a wave function captures quantum correlations at the ``exact exchange" level.   Therefore, as long as correlations due to the electron-electron interaction are not too strong, the KS framework provides a very useful first approximation to study the effects of system inhomogeneities on spin entanglement.  For the KS wave function the two-particle density matrix [Eq.~(\ref{rho2})]  can be factored into a product of one-particle density matrices. 
\begin{figure}[h!]
  \centering
    \includegraphics[width=0.45\textwidth]{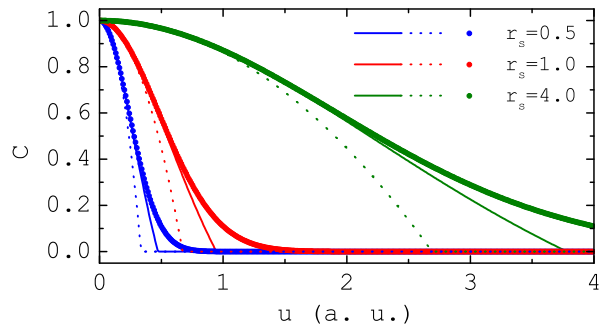}
      \caption{(Color online) Solid lines represent the concurrence of the Fermi gas for different values (represented with different colors) of the Wigner-Seitz ($r_s = 0.5, 1.0, 0.4$ a.u.). 
       Dashed lines and bullets represent the approximate expressions of the concurrence given in Eq.~(\ref{TC}) and Eq.~(\ref{CNL}), respectively. }\label{fig1}
\end{figure}
It is then easily checked that, for a closed-shell system (equal numbers of up and down spins), 
the quantity in Eq.~(\ref{werner}) acquires the form
\begin{eqnarray}\label{rho2ks}
\rho_{2}(\vr_1,\vr_2) & = &  
n(\vr_1)  \left[ n(\vr_2) +  h_{\rm X}(\vr_1,\vr_2) \right] {\mathcal I}/4 \nonumber \\
& - &  n(\vr_1) h_{\rm X}(\vr_1,\vr_2) | \Psi^{-} \rangle \langle \Psi^{-} |/2,
\end{eqnarray}
where $n(\vr)$ is the electron density, $h_{\rm X}(\vr_1,\vr_2)$ is the exchange-hole function
\be\label{hx}
h_{\rm X}(\vr_1,\vr_2) = - | \rho_{1}(\vr_1,\vr_2) |^2 / n(\vr_1)\,,
\ee
and $\rho_{1}(\vr_1,\vr_2)$ is the KS one-body reduced density matrix. This is 
expressed in terms of the (occupied) Kohn-Sham orbitals $\varphi_{i}$ as follows:
\be\label{P1b}
\rho_{1}(\vr_1,\vr_2) =  2~\sum_{i=1}^{N_{\rm occ}} \varphi_{i }(\vr_1)\varphi^*_{i}(\vr_2)\,.
\ee

The exchange-hole function quantifies the correlation between two electrons with equally oriented spins dictated by their fermionic
character. In particular, such correlation may result in an excess of the singlet-state occupation in the two-body density matrix ($p>0$), 
and, possibly, in two-particle spin entanglement ($p>1/3$).
Comparison between Eq.~(\ref{werner}) and Eq.~(\ref{rho2ks}) allows one to express the probability $p$ in terms of the exchange-hole function:
\be\label{pr1r2}
p(\vr_1,\vr_2) =-
\frac{h_{\rm X}(\vr_1,\vr_2)}{2 n(\vr_2)+h_{\rm X}(\vr_1,\vr_2)} .
\ee
The condition for the existence of spin entanglement [$C>0$, see Eq.~(\ref{conc})] is correspondingly given by:
\be\label{SCX}
 h_{\rm X}(\vr_1,\vr_2) < - n(\vr_2)/2\;.
\ee

In order to get further insight, let us consider the case in which only one occupied orbital is significantly different from zero at both $\vr_1$ and $\vr_2$.
In this case, $h_{\rm X}(\vr_1,\vr_2) \simeq - n(\vr_2) $, and thus $C(\vr_1,\vr_2) \simeq 1$.
This applies, approximately, in the asymptotic region of any atom or molecule, and in the regions of the atomic shells and molecular bonds.   As $\vr_2$ moves away from a fixed $\vr_1$, the concurrence may thus have revivals  even at distances comparable to the system size, if $\vr_2$ is positioned within the same shell or bond as $\vr_1$. This is completely different from the monotonic decrease of $C$ that one expects in a uniform Fermi gas.

{\it Short-range behavior of the concurrence --}
The example of the non-interacting Fermi gas suggests that, locally the concurrence should decrease with some characteristic length.
In order to investigate this important aspect in realistic many-electron systems, 
we introduce the spherical average of the two-body density matrix [Eq.~(\ref{rho2ks})]:
\be
\overline\rho_{ 2}(\vr,u) = \frac{1}{4\pi} \int d \Omega_u~ \rho_{2}(\vr_1 = \vr,\vr_2 = \vr + \vu)\,,
\ee
where $\Omega_u$ is the solid angle defined by $\vu$ around $\vr$.
Analogous averages apply to the exchange-hole and to the particle density: they
are referred hereafter as $ \overline h_{\rm X} (\vr , u) $ and $ \overline n(\vr , u) $, respectively.
The short-range behavior of these quantities can be derived from their Taylor expansion in the interparticle distance $u$~\cite{Becke83,Dobson93}:
\begin{eqnarray}
\label{Trho}
\overline n (\vr,u) \simeq n(\vr) + \frac{1}{6} \nabla^2 n(\vr) u^2 + \cdot\cdot\cdot
\\
\label{ThX}
\overline h_{\rm X}(\vr,u) \simeq - n(\vr) + \frac{1}{3} D(\vr) u^2  + \cdot\cdot\cdot\;,
\end{eqnarray}
where
\be\label{D}
D(\vr) = \tau(\vr) 
-\frac{1}{4} \frac{\left[ \nabla n(\vr) \right]^2}{n(\vr)} 
- \frac{\vj_{p}^{\,2}(\vr)}{n(\vr)}\;.
\ee 
In the above equation, 
\be
\tau(\vr) = 2~\sum_{i = 1}^{N_{\rm occ}} | \nabla \varphi_{i}(\vr) |^2\;
\ee
is (twice) the positive definite kinetic energy density, and 
\be
\vj_{p}(\vr) =  \frac{1}{i} \sum_{i=1}^{N_{\rm occ}} \left[ \varphi^{*}_{i}(\vr) \nabla \varphi_{i}(\vr) - \varphi_{i}(\vr) \nabla \varphi^{*}_{i}(\vr) \right] \;
\ee
is the KS paramagnetic current density.
Note that $\vj_{p}  = 0$ in the ground-state of closed-shell systems, but can  very well be different from zero in a time-dependent situation.

The short-range (SR) behavior of the concurrence is readily derived by replacing the
Taylor expansions of  $h_{\rm X}$ and $n$ given above in Eqs.~(\ref{conc},\ref{pr1r2}).
The resulting expression is given, to the lowest order in $u$, by
\be\label{TC}
C_{\rm SR}(\vr,u) = \max \left\{ 0, 1 - u^2 D(\vr)/n(\vr) \right\}.
\ee
Equation \ref{TC} naturally leads to the introduction of a length scale, 
$ l_E(\vr) \equiv [D(\vr) / n(\vr) ]^{-1/2} $, 
which expresses the  ``local range" of the spin entanglement around a given point in space. 
Combining such expression with Eq. (\ref{D}), the length scale reads:
\be\label{lE}
l_{E} (\vr) \equiv \left\{ \frac{\tau(\vr)}{ n(\vr)}
-\frac{1}{4} \frac{\left[ \nabla n(\vr) \right]^2}{[ n(\vr)]^2} 
- \frac{\vj_{p}^2(\vr)}{[ n(\vr)]^2} \right\}^{-1/2} .
\ee

\begin{figure}[h!]
  \centering
    \includegraphics[width=0.45\textwidth]{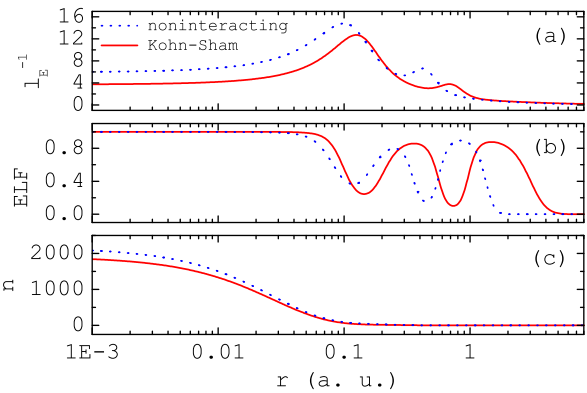}
      \caption{(Color online) Argon atom. Radial dependences of the inverse of the local entanglement-length (a),  of the ELF (b), and of the particle density  $n$ (c).
      All the input quantities have been obtained using the APE code~\cite{APE}.
      The approximation   to the exchange-correlation energy functional employes Dirac exchange~\cite{ldax} and Perdew 
       and Zunger correlation~\cite{ldac}.}\label{fig2}
\end{figure}
It is tempting to extrapolate the behavior of the concurrence to larger interparticle distances, through
an (approximate) exponential resummation,
as follows
\be\label{CNL}
C(\vr,u) =  \exp \left[ - u^2 / \, l^{2}_{E}(\vr) \right]\;.
\ee
Figure~{\ref{fig1}} compares the concurrence of the noninteracting Fermi gas, as a function of $u$, with the one obtained from the uncontrolled extrapolation of the small-$u$ expansion [Eq.~(\ref{TC})], and from the more educated extrapolation [Eq.~(\ref{CNL})].  This is done for values of the Wigner-Seitz radius $r_s$ ranging from typical metallic densities to higher ones. Readily, one finds that $l_E^{\rm unif} = \sqrt{\frac{5}{3}} \frac{1}{ k_{\rm F}} $ (with $ k_{\rm F} = \left( 3 \pi^2  n\right)^{1/3}$).
As expected, the higher the density, the more the entanglement is short-ranged, the more Eq.~(\ref{TC}) gets accurate. 
Equation~(\ref{CNL}) tends to recover the exact concurrence also at intermediate interparticle distances, although it introduces a spurious tail for larger values of $u$.
More importantly, Eq.~(\ref{TC}) and Eq.~(\ref{CNL}) {\it also} apply -- within the specified limitations and approximations -- to non-uniform gases. 
Consideration of the latter cases brings us to find a useful connection between the entanglement length and the so-called electron localization function.

{\it Entanglement length and electron localization function --}
The electron localization in the proximity of a reference position for a given spin is (inversely) related to the conditional probability of finding a spin-like electron~\cite{BE90, BMG05}.
This brings to define the so-called {\it electron localization function}:
\be\label{ELF}
ELF(\vr) = \frac{1}{1 +   \chi^2(\vr) }, 
\ee
where $ \chi(\vr) \equiv D(\vr) / D^{\rm unif}(\vr) $ 
and 
$ D^{\rm unif}(\vr) = \frac{3}{10}(3\pi^2)^{2/3} [n (\vr)]^{5/3}$. Empirically, Eq.~(\ref{ELF}) was found to effectively visualize atomic shells and molecular bonds. 
Since then, the ELF has became a valuable tool in understanding the chemical structure of molecules and materials~\cite{Savin92}.
Here we show that, quite remarkably, the ELF can be rewritten in terms of the entanglement length:
\be\label{ELF2}
ELF(\vr) = \frac{1}{1 +   \left[ l^{\rm unif}_E(\vr)/l_E(\vr) \right]^2 }\;.
\ee
Therefore, the ELF quantifies the difference between the actual entanglement length  and the one of a uniform gas having the same particle density at the position of interest. 
Next, we consider in detail the case of an atom and of a simple molecule.

{\it Atomic systems --}
 As a representative example of a closed-shell atom, we consider the case of Ar.
Panel (a) of Fig.~(\ref{fig2}) shows the {\it inverse} of $l_E(\vr)$ rather than the length itself, for the sake of a simpler visualization.
This quantity exhibits a strong dependence on the shell structure. 
In particular, the local maxima of $l^{-1}_E (\vr) $ correspond to local minima of the ELF [panel (b)], and viceversa.
Therefore, the entanglement gets more short-ranged between atomic shells and  more long-ranged within the atomic shells.
Moving outwardly from the last shell, one enters  the asymptotic region of the Ar atom. Here, the entanglement length tends to infinity --
but the probability of finding an electron vanishes exponentially for increasing $r$. 
As anticipated, the ELF can be reinterpreted as an indicator of the difference between $l_E (\vr)$ and the entanglement length of a uniform gas having the same particle density as the original system at  point $\vr$.  Note that $l^{\rm unif}_E (\vr)$ can only follow the structureless profile of the particle density [panel (c)], 
and thus fails to capture features such as shells and bonds.
Therefore, the structures in the ELF reflects the effects of the system inhomogeneities in the spatial distribution of entanglement. 

It is worth asking to what extent the Coulomb interaction between electrons affects the distribution of entanglement. 
We may get some insight by comparing the previous results with those obtained by setting the Hartree and exchange-correlation potentials to zero in the Kohn-Sham equation (dashed curves in Fig.~\ref{fig2}).
Underlying this comparison is the idea that the single-particle orbitals and eigenvalues of the Kohn-Sham 
systems represent, in a first approximation, quasi-particles obtained 
by screening the otherwise purely non-interacting quantities with correlations of both classical (Hartree) and quantum mechanical (exchange-correlation) origin.
The most prominent difference between the KS and the truly non-interacting solution is that, in the latter, both the shell structure and particle density move towards the core of the atom. Besides, the local entanglement-length is reduced both in correspondence of the atomic shells and in the spatial region between them.  

\begin{figure}[h!]
  \centering
    \includegraphics[width=0.48\textwidth]{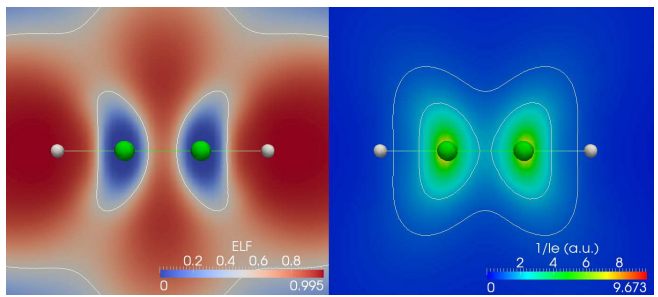}
      \caption{(Color online) $C_2H_2$ molecule.  Left panel shows the ELF in the $xy$ plane.
      Right panel shows the inverse of the local entanglement-length, $l^{-1}_E (\vr)$.
       Also note the different scales used in the two panels. All the input quantities have been obtained using the OCTOPUS code~\cite{OCTOPUS}. 
       The approximation  to the exchange-correlation energy functional employes Dirac exchange~\cite{ldax} and Perdew 
       and Zunger correlation~\cite{ldac}.}\label{fig3}
\end{figure}

{\it Molecular systems --- }
The above results suggest that the SR behavior of entanglement in molecules might be affected by the presence of covalent bonds. To this aim, we consider the case of C$_2$H$_2$. Being mainly interested in the bond region, we perform our calculation with pseudopotentials for the Carbon atoms, and explicitly include only the $2s$ and $2p$ electrons.
The spatial dependence of the ELF and $l^{-1}_E (\vr)$ are reported in Fig.~\ref{fig3} (left and right panels, respectively).  
The ELF displays a typical squeezed toroidal structure in correspondence of the central bond between the two Carbon atoms, where $l^{-1}_E (\vr)$ has a local minimum. Therefore, entanglement gets relatively long-ranged not only in the atomic shells, but also within the bond region. The plot of $l^{-1}_E (\vr)$ also reveals that entanglement is long-ranged in the asymptotic region of the molecule and around each Hydrogen atom, where 
the two-spin state essentially coincides with a singlet. These results confirm that --  in the case of molecules as well -- the ELF fully captures the effects of the system
inhomogeneities on the spatial distribution of entanglement.

{\it Conclusions ---}  
The results and ideas presented in this Letter deepen our understanding of 
spin entanglement in the ground state of many-electron systems from an {\it ab-initio} perspective.
We have shown that spin entanglement in atomic and molecular systems behaves in ways that are strongly affected by the non-homogeneity of the system.
The entanglement length has maxima in correspondence of atomic shells and molecular covalent bonds. 
The electron localization function has been re-expressed in terms of the 
ratio between the entanglement length of the system and that of an homogeneous gas having locally the same particle density.
Therefore the ELF provides a useful estimate of the effects of system inhomogeneity on  spin entanglement.
We have also shown how to capture the nonlocal functional dependence of the concurrence on the particle density,  by including the  KS orbitals explicitly in the description.  An extension of our analysis to time-dependent states, within the framework of time-dependent DFT, is straightforward.  
A more challenging proposition will be the generalization of the present approach to systems in which a more accurate treatment of Coulomb correlations beyond the exchange level  is required.

\begin{acknowledgments}
{\it Acknowledgments --} This work was financially supported by the European Community through the FP7's CRONOS project, grant agreement
no. 280879 (S.P. and C.A.R.); the Italian FIRB Project No. RBFR12RPD1 of the Italian MIUR (F.T.); and DOE-BES Grant No. DE-FG02-05ER46203 (GV).
\end{acknowledgments}

\bibliography{paper}

\end{document}